\documentclass[12pt]{article}
\usepackage{times}
\usepackage{geometry}
\usepackage[utf8]{inputenc}
\usepackage{enumitem,amssymb}
\usepackage{ragged2e}
\newlist{thematic}{itemize}{8}
\setlist[thematic]{label=$\square$}
\usepackage{pifont}

\usepackage{graphicx}
\usepackage{amsmath}%
\usepackage{amsfonts}%
\usepackage{amssymb}%
\usepackage{subfig}
\usepackage{wrapfig}
\usepackage{scrextend}
\usepackage{lineno}
\usepackage[sort&compress,numbers]{natbib}
\usepackage{enumitem}
\usepackage{fancyhdr}
\usepackage{xcolor}
\usepackage{ifthen}
\usepackage{ragged2e}
\usepackage{multicol}

\begin{document}
\raggedright
\huge
Astro2020 Science White Paper \linebreak

Prospects for the detection of synchrotron halos around middle-age pulsars \linebreak
\normalsize

Multi-Messenger Astronomy and Astrophysics \hspace*{65pt} \linebreak
  
\textbf{Principal Author:}

Name: Mattia Di Mauro
 \linebreak						
Institution:  NASA Goddard Space Flight Center and Catholic University of America, Department of Physics
 \linebreak
Email: mdimauro@slac.stanford.edu
 \linebreak	
 
\textbf{Co-authors:}  
  \linebreak
Name: Manconi Silvia, Donato Fiorenza
 \linebreak						
Institution:  Dipartimento di Fisica, Universit\`a degli Studi di Torino and Istituto Nazionale di Fisica Nucleare, Italy \linebreak
Email: manconi@to.infn.it, donato@to.infn.it
\linebreak

\textbf{Endorsers:}  
  \linebreak
Hui Li, Los Alamos National Laboratory, USA
  \linebreak
Silvia Zane, Mullard Space Science Laboratory, University College London, UK
\linebreak

\textbf{Abstract  (optional):}
The origin of cosmic-ray positrons detected with an energy above 10 GeV is one of the most intriguing mysteries in Astroparticle Physics.
Different interpretations have been invoked to solve this puzzle such as pulsar wind nebulae (PWNe), supernovae remnants or interactions of dark matter particles.
The HAWC and Milagro experiments have measured an extended emission of photons above 10 TeV in the direction of Geminga and Monogem PWNe which can be interpreted by photons emitted through inverse Compton scattering by positrons and electrons accelerated by these sources.
These HAWC and Milagro detections are extremely important to estimate the PWN contribution to the positron excess.
Positrons and electrons emitted by PWNe and injected in the interstellar medium can produce photons with an energy between radio and X-ray through synchrotron radiation with the Galactic magnetic field.
In this white paper we show that Astrophysics Probe mission concepts, such as AMEGO (the All-sky Medium Energy Gamma-ray Observatory) and AdEPT (The Advanced Energetic Pair Telescope), are suitable to detect the synchrotron halos from Monogem and Geminga PWNe and we will report the list of the most promising PWNe that AMEGO could be able to detect.

\pagebreak

\section{Introduction}
\vspace{-3mm}
Satellite experiments such as AMS-02, Pamela, {\it Fermi}-LAT, DAMPE and CALET and the Imaging Atmospheric Cherenkov Telescopes (IACTs) HESS and VERITAS have published data for the spectrum and anisotropy of cosmic-ray (CR) positrons ($e^+$) and electrons ($e^-$) \cite{2009Natur.458..607A,2012PhRvL.108a1103A,PhysRevLett.113.121101,PhysRevLett.120.261102,Ambrosi:2017wek,PhysRevLett.122.041102}.
The data for the inclusive flux ($e^{+}+e^-$) cover many orders of magnitude (from 0.1 GeV to 20 TeV) and have reached a precision as low as a few \%.
This incredibly rich dataset could be used to determine the emission mechanism of CRs from Galactic sources such as pulsar wind nebulae (PWNe) and supernova remnants (SNRs), to model the CR propagation in the Galaxy and to find signatures of exotic contributions from, for example, dark matter (DM) particle interactions.

The origin of the $e^+$ flux and the positron fraction ($e^+/(e^+e^-)$) remains still one of the most intriguing mysteries that physicists are struggling to solve since about a decade.
Below 10 GeV the $e^+$ flux is well described with the secondary mechanism given by spallation reactions of primary 
CRs with the atoms of the interstellar medium (ISM) (see, e.g., \cite{2014JCAP...04..006D}).
On the other hand, the $e^+$ flux above a few tens of GeV strongly exceeds the predicted secondary component. 
This is the so-called $e^+$ excess which has been interpreted with the annihilation or decay of DM particles (see, e.g., \cite{2016JCAP...05..031D}) or the the emission from PWNe (see, e.g.,  \cite{2014JCAP...04..006D}) or SNRs (see, e.g., \cite{2009PhRvL.103e1104B}).

The Milagro and more recently the HAWC Collaborations have detected above 10 TeV an extended emission of $\gamma$ rays in the direction of Geminga pulsar with an angular size of the order of $2^{\circ}$ \cite{2009ApJ...700L.127A,Abeysekara:2017science}. 
The HAWC experiment has also detected the same feature around Monogem pulsar and other Galactic sources.

These measurements are of central importance in the understanding of the contribution of PWNe to the $e^+$ excess.
The $e^\pm$ pairs accelerated and emitted by the PWNe suffer severe energy losses, which in turn give origin to a cascade of photons in a broad range of frequencies.
In particular, the very-high energy (VHE) $\gamma$-ray emission can be explained as photons produced through $e^{\pm}$ emitted from these sources and inverse Compton scattering (ICS) off the interstellar radiation field (ISRF).
Therefore, the HAWC and Milagro measurements are an indirect probe of the $e^{\pm}$ CRs emitted by Galactic PWNe.

$e^{\pm}$ CRs can also produce photons at lower energy from radio to X-ray through synchrotron radiation on the Galactic magnetic field. This emission should be detectable as an halo of photons in the direction of PWNe similarly to $\gamma$-ray halos detected by Milagro and HAWC.
As we will show in this white paper, the detection of these synchrotron halos is extremely challenging for radio and X-ray telescopes. Indeed, the extension of halos for sources located within a few kpc from the Earth is of the order of the degree which is approximatively the field of view of these experiments.

AMEGO, the All-sky Medium Energy Gamma-ray Observatory, and AdEPT (The Advanced Energetic Pair Telescope) \cite{Hunter:2013wla} are Astrophysics Probe mission concepts designed to explore the sky in the energy range between $0.2-1000$ MeV and 5-200 MeV, respectively.
These observatories are designed to have at 100 MeV a spatial resolution of about $1^{\circ}$ and a sensitivity of about $10^{-6}$ MeV/cm$^2$/s for AMEGO and about three time worse for AdEPT.
Therefore, AMEGO, AdEPT or any other future experiments with a similar design would be ideal for detecting synchrotron halos from middle-age pulsars with an age $>30$ kyr because at MeV energies their extensions, for sources located within a few kpc from the Earth, is expected to be at the degree scale.
In this document we will demonstrate that these two mission concepts are suitable to detect the synchrotron halos from Monogem and Geminga PWNe and we will report the list of the most promising PWNe that they could be able to detect.

\vspace{-6mm}
\section{Photon and $e^{\pm}$ emission from Pulsar Wind Nebulae}
\label{sec:mw} 
\vspace{-3mm}
PWNe are among the major accelerators of $e^+$ and $e^-$ in the Galaxy.  
A PWN is generated under the influence of strong magnetic fields located on the surface of the neutron star which initiate cascade processes and lead to the production of a cloud of charged particles that surrounds the pulsar (see, e.g., \cite{Amato:2013fua,2017hsn..book.2159S}).
This process is thought to accelerate $e^+$ and $e^-$ to VHE which are then injected into the ISM after a few tens of kyr \cite{Amato:2013fua}.
The photon emission observed in the direction of PWNe covers a wide range of energies.
From radio to X-ray energies, photons are produced by $e^+$ and $e^-$  through synchrotron radiation caused by the magnetic field present in the ISM.
On the other hand, at higher energies $\gamma$ rays are produced  via ICS of VHE $e^+$ and $e^-$ escaped from the PWN off the ISRF.

We will assume that $e^{\pm}$ are emitted from the PWNe continuosly from the age of the supernova explosion and with a rate that is proportional to the magnetic dipole braking $L(t)$ ($L(t) = L_0 \left( 1+ t/\tau_0 \right)^{(k+1)/(k-1)}$, where $k$ is the magnetic braking index, taken typically to be $k=3$, and $\tau_0$ is the typical pulsar decay time. We will assume here $\tau_0=12$ kyr.
In literature the burst like scenario has been considered too (see, e.g., \cite{2014JCAP...04..006D}). In this model all the particles are emitted from the sources at a time equal to the age of the source ($t^{\star}$).
However, in this model it is not possible for a PWN like Geminga with an age $t^{\star}=342$~kyr and a distance $d=0.25$ kpc to produce through ICS $\gamma$ rays above 5 TeV, as detected by HAWC and Milagro.
Indeed, considering energy losses given by $dE/dt=b(E)=10^{-16} (E/1\rm{GeV})^2$ GeV s$^{-1}$, the maximum energy of an electron emitted by the PWN at a time $t^{\star}$ is about $0.9$ TeV.

We calculate the flux of $e^\pm$ as a function of energy, position in the Galaxy, and time as in \cite{2009PhRvL.103e1101Y} while the photon flux emitted for ICS or synchrotron mechanism by a source as in \cite{Cirelli:2010xx}.
Finally, we define the ICS power of photons of energy $E_{\gamma}$ produced by electrons of energy $E$ as in \cite{2010A&A...524A..51D} while the synchrotron power is defined as in \cite{2010PhRvD..82d3002A}.

\vspace{-6mm}
\section{Science case}
\begin{figure}
\centering
\includegraphics[width=3.2in]{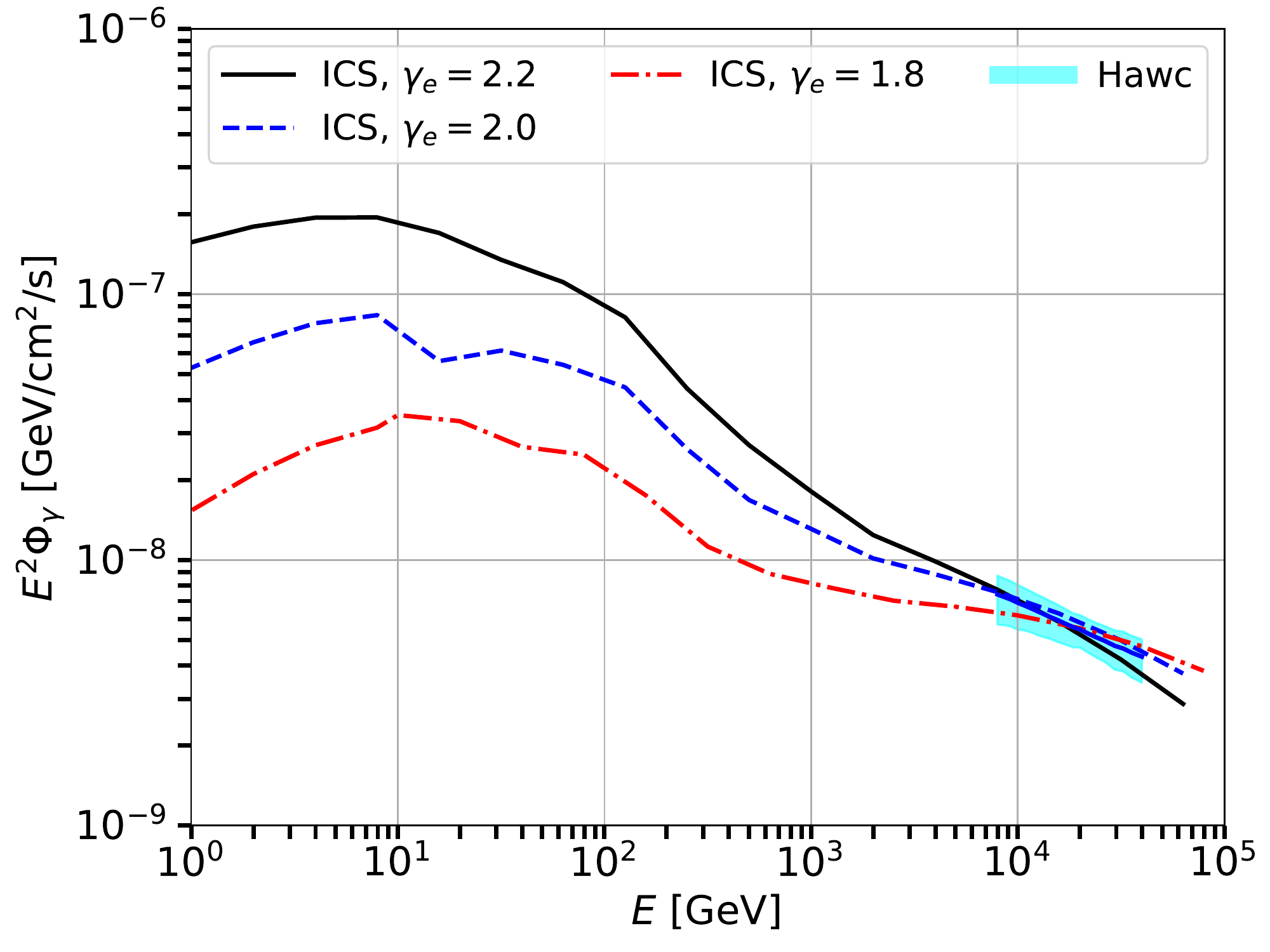}
\includegraphics[width=3.2in]{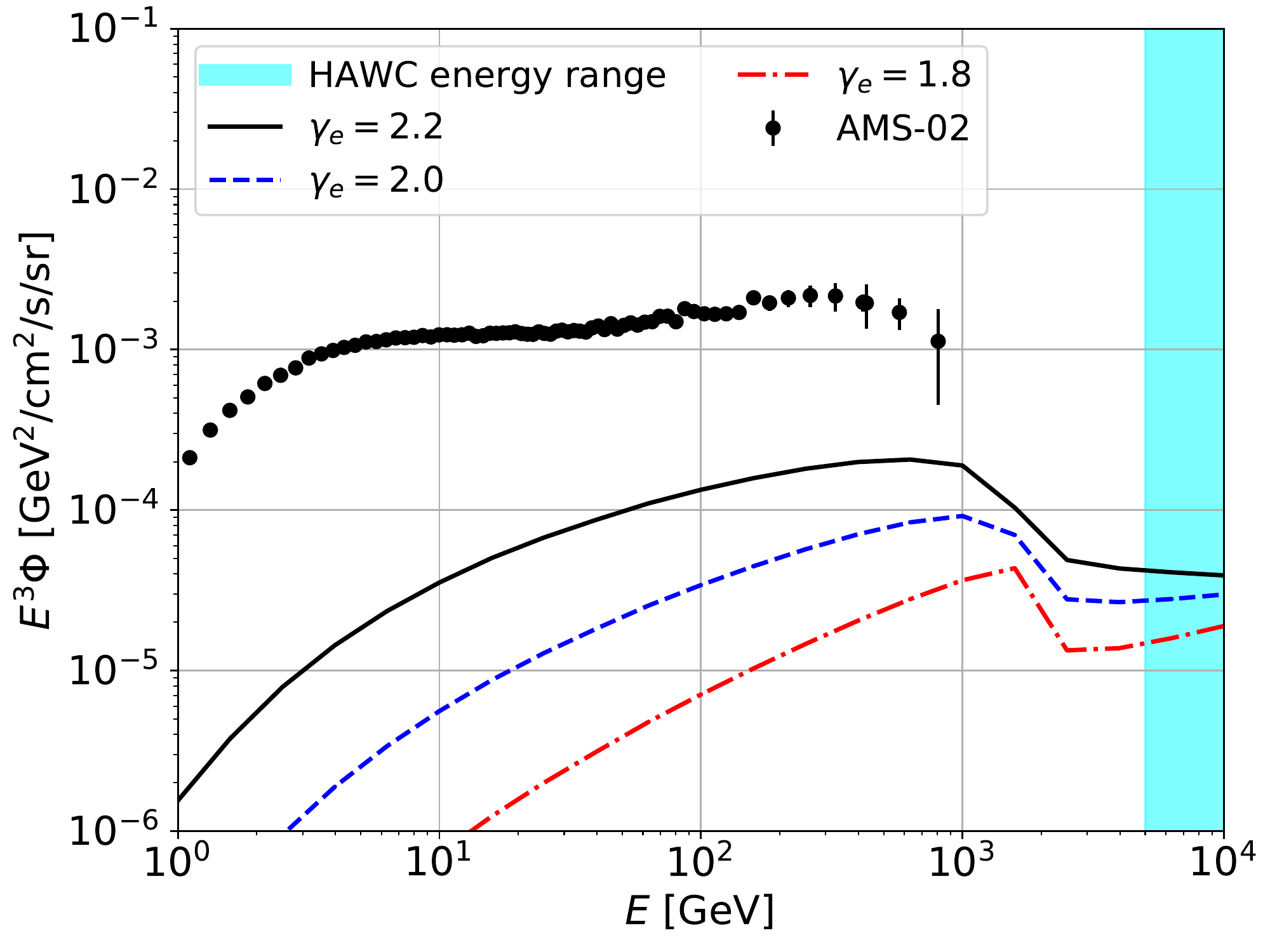}
\caption{ {\footnotesize Left panel: $\gamma$-ray flux $\Phi_{\gamma}$ for Geminga ICS halo for three different choices of the positron spectral index $\gamma_{e}$ compared to HAWC measurements (cyan band). Right panel: positron flux from Geminga PWN compared to AMS-02 data (black data).} \label{fig:GemingaICS}}
\end{figure}

\vspace{-3mm}
We consider Geminga here as a benchmark to illustrate the potential for AMEGO to observe and constrain the $e^\pm$ emission from PWNe.
The HAWC measurements for Geminga and Monogem ICS halos are at $\gamma$-ray energies between $5-50$ TeV.
In the left panel of Fig.~\ref{fig:GemingaICS} we show the predictions for the ICS $\gamma$-ray flux calculated for Geminga PWN in comparison with HAWC data. We perform this calculation for three different choices of the electron spectral index: $\gamma_e=[1.8,2.0,2.2]$.
The normalization of the $e^{\pm}$ injections spectrum ($Q(E)$) is chosen in order to produce an ICS flux compatible with HAWC data. 
The normalization of $Q(E)$ translates into an efficiency for the conversion of spin-down energy into $e^\pm$ pairs between $1-2\%$ depending on the choice of $\gamma_e$ values used here.
It is important to notice that the photon flux extrapolated at $E_{\gamma}<10^2$ GeV can vary by one order of magnitude.
In the right panel of the same figure we show the correspondent predicted $e^{+}$ flux at Earth for the same three values of $\gamma_e$ considered before.
In Fig.~\ref{fig:GemingaICS} (right panel) we show the predicted $e^+$ flux at Earth for each value of $\gamma_e$ as considered before.
They provide similar fluxes above 10 TeV, for which $\gamma$ rays are detected by HAWC, while in the energy range of AMS-02 data their $e^+$ flux varies significantly.
Geminga can contribute to the $e^+$ AMS-02 high-energy data at most $20\%$ for $\gamma_e=2.2$, and $5\%$ with $\gamma_e=1.8$ in the highest energy AMS-02 data point.
The main consequence of this analysis is that with HAWC data alone it is not possible to constrain precisely the contribution of Geminga and Monogem PWNe to the AMS-02 $e^+$ excess. 

This result is not unexpected, since the $e^+$ flux is tuned to HAWC data at $E_\gamma>5$ TeV. 
In the Thomson regime, a $\gamma$ ray detected at 10 TeV is on average produced by a $e^+$ or an $e^-$ at energies around 60 TeV through ICS with the CMB. During the propagation from the source to Earth $e^+$ lose energy through synchrotron radiation and ICS and $e^+$ are detected with an energy of about 2 TeV\footnote{We assume for this simple calculation that the energy of a photon produced by a positron with a Lorentz factor of $\gamma$ for ICS on a photon field given by a black body distribution with a characteristic temperature of $T_{{\rm BB}}$ is $3.60 \gamma^2 k_B T_{{\rm BB}}$. Moreover we assumed energy losses given by $5\cdot 10^{-17} E^2$ GeV/s that are compatible with a Galactic magnetic field with $3\mu$G and the ISRF of \cite{Vernetto:2016alq,Porter:2005qx}.}.
Therefore, $\gamma$-ray data between $1-100$ GeV from {\it Fermi}-LAT are highly desirable in order to constrain more precisely the Monogem and Geminga contribution to the $e^+$ excess.

However, the detection of the ICS emission from PWNe in this energy range is extremely challenging because of the contribution of the Galactic diffuse emission that below $10^2$ GeV becomes by far the dominant component.
Moreover, below 10 GeV the pulsed emission of the pulsar emerges as a bright component.
Indeed, Geminga pulsar above 10 GeV is detected at about 120 $\sigma$ significance and is among the brightest sources detected in the $\gamma$-ray sky.
The most relevant reason that makes challenging the detection of these ICS halos in the {\it Fermi}-LAT energy range is that the halo is very extended. 
In the case of Geminga for example with a diffusion coefficient of $D_0 = 7\cdot 10^{25} {\rm cm}^2/s$ and $\delta=0.33$ the ICS halos has an extension of about $15^{\circ}$ at 10 GeV and $8^{\circ}$ at 100 GeV.
It is very difficult to disentangle between the isotropic emission or large scale structures of the Galactic diffuse emission with such an extended halo.
Finally, the Geminga pulsar has a proper motion of about $211$ km/s that makes the pulsar move by about 70 pc across its age. Therefore, a proper analysis to detect the ICS halo from this source should include the pulsar proper motion.

\begin{figure}
\centering
\includegraphics[width=3.2in]{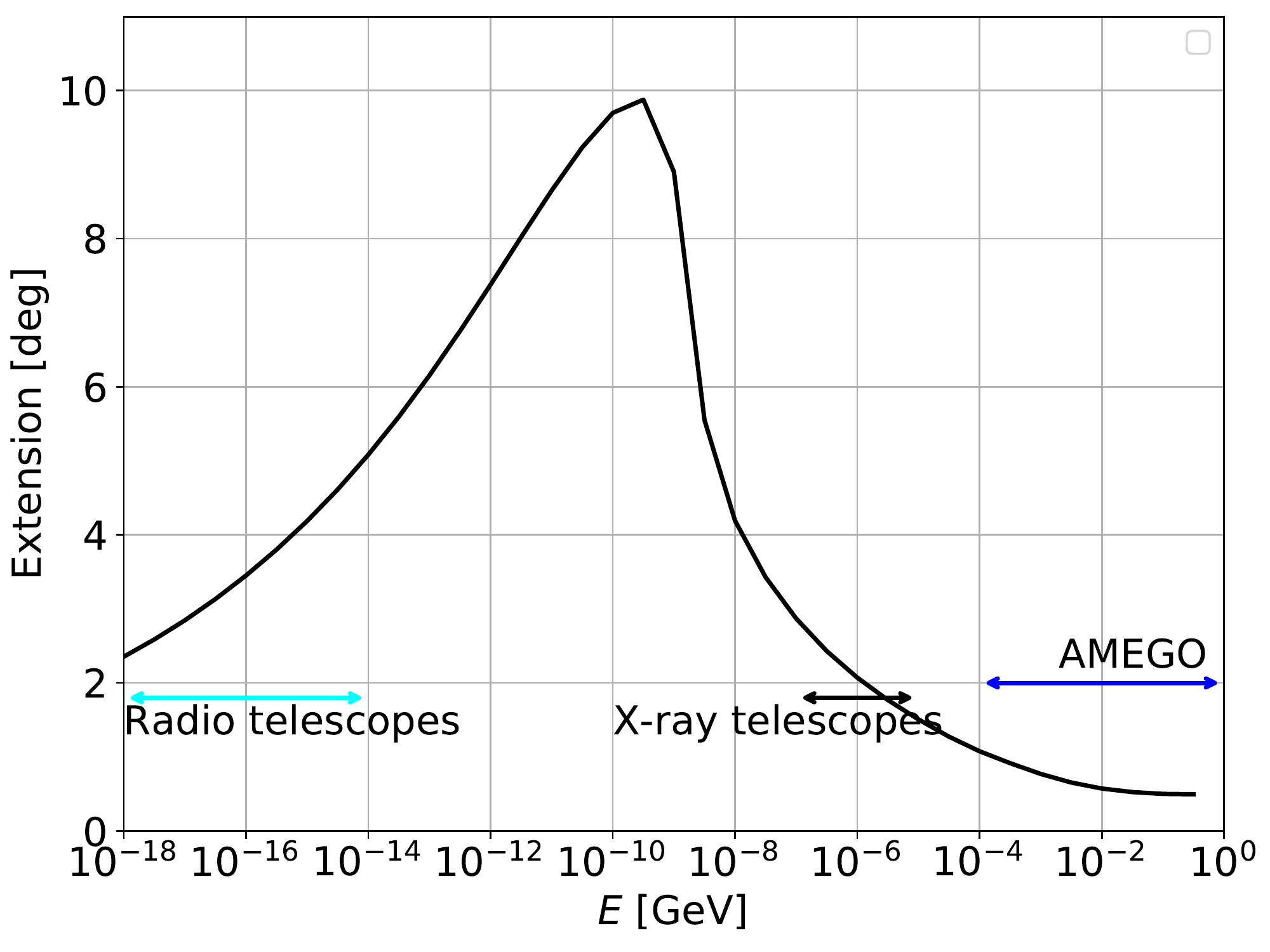}
\includegraphics[width=3.2in]{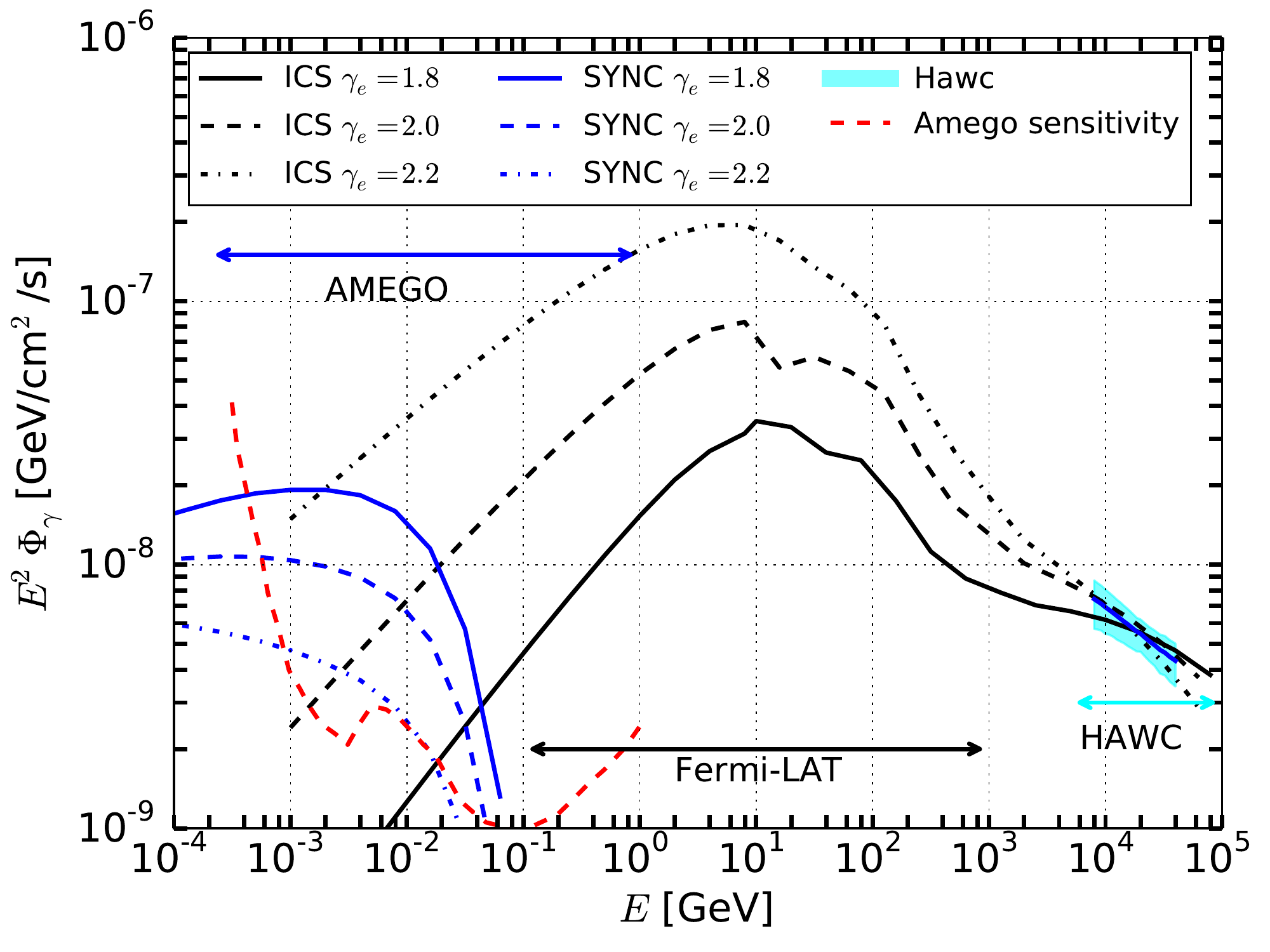}
\vspace{-3mm}
\caption{{\footnotesize Left panel: Angular extension of the Geminga PWN synchrotron halo between radio up to $\gamma$-ray energies. Right panel: Geminga PWN photon flux for ICS (black lines) and for Synchrotron radiation (blue lines) calculated for three different choices of $\gamma_e$. We also diplay the $3\sigma$ continuum sensitivity for Amego (red dashed line).}}
\label{fig:GemingaAMEGO}
\end{figure}


As explained above, $e^{\pm}$ emitted from PWNe produce photons by synchrotron radiation from radio to $\gamma$-ray energies.
$\gamma$ rays in the HAWC energy range are emitted by $e^{\pm}$ with an energy of about 60 TeV and these $e^{\pm}$ in turns produce photons for synchrotron radiation at a typical energy of a few keV.
However, in the continuous injection scenario, electrons could be produced at energies much higher than 60 TeV and photons are created for synchrotron radiation even at MeV energies, if no cutoff is present in the PWN injection spectrum below 100 TeV. Therefore, in the following we use a cutoff energy value of $10^{4}$ TeV.
We also stress that for gyroscale acceleration in radiation reaction, the maximum synchrotron energy is around 160 MeV \cite{1996ApJ...457..253D}.

Radio and X-ray telescopes are possible experiments which can be used to search for these synchrotron halos.
They cover exactly the energy range of this emission.
However, at radio energies and in the case of Geminga, the halo would be extended about $4^{\circ}$ while in the energy range covered by X-ray telescopes ($0.1-10$ keV) its size is $2^{\circ}$ as shown in Fig.~\ref{fig:GemingaAMEGO} (left panel).
It is thus prohibitive for radio and X-ray telescopes to detect these halos because they have a field of view smaller or of the same order of them.

The AMEGO mission on the other hand is planned to work in the energy range between $0.2-1000$ MeV.
The extension of the synchrotron halos for Geminga is expected to be of the order of $1^{\circ}$ in this energy range.
The AMEGO field of view which will be 2.5 sr and the spatial resolution of the order of $2^{\circ}$ make this experiment ideal for this scope.

In order to show that this mission concept would be promising to detect synchrotron halos, we calculate the synchrotron flux for the same cases displayed for Geminga in Fig.~\ref{fig:GemingaICS}.
We show the results in Fig.~\ref{fig:GemingaAMEGO} (right panel) together with the simulated $3\sigma$ continuum sensitivity for AMEGO. We see from this figure that AMEGO will probably be able to detect the synchrotron halo from Geminga regardless the value of the positron spectral index.
Moreover, a future detection by AMEGO will help to give more information about the value of $\gamma_e$ thus providing an additional pivot to give more precise predictions for the contribution of this PWN to the $e^+$ excess.
We also find by applying the same method to Monogem PWN that AMEGO should be able to detected the synchrotron halo from this sources if the positron spectral index is harder than 2.2.

As reported before, we assume in our model a cutoff energy in the positron injection spectrum of $10^4$ TeV.
VHE $\gamma$ rays have been detected from PWN even at 100 TeV (see, e.g., \cite{2019A&A...621A.116H}) proving indirectly that $e^{\pm}$ at energies $\geq 10^3$ TeV are probably accelerated and emitted from these objects. However, it is not clear what is the highest possible energy at which PWNe produce $e^{\pm}$. Therefore, the detection by AMEGO of MeV photons emitted by PWNe would be a strong evidence for the nature of these sources as ultra high-energy CR accelerators.

\vspace{-6mm}
\section{Synchrotron halos detectable by AMEGO}
\vspace{-3mm}
\begin{table}[t]
\center
\begin{tabular}{|c|c|c|c|c|c|}
\hline 
Name & l [deg] & b [deg] & age [kyr]    &  dist [kpc]  &   $\dot{E}$ [erg/s]   \\
\hline 
B1055-52 &  285.984 & 6.649 & 535 & 0.09 & $3.01\cdot 10^{34}$ \\
J0633+1746 &  195.134 & 4.266 & 342 & 0.19 & $3.25\cdot 10^{34}$ \\
J1813-1246 &  17.244 & 2.445 & 43 & 2.63 & $6.24\cdot 10^{36}$ \\
B1951+32 &  68.765 & 2.823 & 107 & 3.0 & $3.74\cdot 10^{36}$ \\
J1105-6107 &  290.49 & -0.846 & 63 & 2.36 & $2.48\cdot 10^{36}$ \\
B0656+14 &  201.108 & 8.258 & 111 & 0.29 & $3.81\cdot 10^{34}$ \\
B0906-49 &  270.266 & -1.019 & 112 & 1.0 & $4.92\cdot 10^{35}$ \\
J1809-2332 &  7.39 & -1.995 & 68 & 0.88 & $4.3\cdot 10^{35}$ \\
J1044-5737 & 286.574 & 1.163 & 40 & 1.9 & $8.03\cdot 10^{35}$ \\
J1112-6103 &   291.221 & -0.462 & 33 & 4.5 & $4.53\cdot 10^{36}$ \\
J1459-6053 & 317.886 & -1.791 & 65 & 1.84 & $9.09\cdot 10^{35}$ \\
J1954+2836 &  65.244 & 0.377 & 69 & 1.96 & $1.05\cdot 10^{36}$ \\
J1524-5625 & 323.0 & 0.351 & 32 & 3.38      & $3.21\cdot 10^{36}$ \\
J1732-3131 & 356.307 & 1.007 & 111 & 0.64 & $1.46\cdot 10^{35}$ \\
J1028-5819 & 285.065 & -0.496 & 90 & 1.42 & $8.32\cdot 10^{35}$ \\
\hline 
\end{tabular}
\caption{Characteristics of the brightest middle-age pulsars at $E_{\gamma}=3$ MeV in the ATNF catalog.}
\label{tab:detectable}
\end{table}

We have presented in the previous section that AMEGO would be able to detect the synchrotron halo in the direction of Geminga and Monogem pulsars confirming the discoveries of ICS halos made by MILAGRO and HAWC.
Geminga is one of the brightest pulsar in $\gamma$ rays and many others are probably bright enough to be detected by AMEGO.
In order to estimate the number of detectable synchrotron halos we take the publicly available ATNF catalog\footnote{We use the continuously updated list version 1.59 located here \url{https://www.atnf.csiro.au/people/pulsar/psrcat/}.}.
We calculate the synchrotron flux for all pulsars with an age larger than 30 kyr, since we are interested in middle-age pulsars, and we rank their flux at the photon energy of 3 MeV.
We choose this energy because at higher energy the pulsar would have to accelerate an even more energetic electron at more than $10^{2}-10^{3}$ TeV and as we discussed before this is physically challenging for the Physics involved in PWNe. Secondly, at higher energy we have the contamination of the ICS halo and the pulsed emission. 
Among the brightest pulsars expected to produce bright synchrotron halos we have Monogem (B0656+14) Geminga (J0633+1746), J1813-1246, B1951+32, J1105-6107, B0906-49, J1809-2332, J1044-5737, J1112-6103, J1459-6053.
Considering an efficiency for the conversion of energy into electrons and positrons pairs of $1\%$ we find that AMEGO should be able to detect about 30 synchrotron halos from middle-age pulsars.

These detections would be extremely important for the study of CR $e^{\pm}$ production mechanisms from PWNe, for the interpretation of the AMS-02 $e^+$ excess and for opening a new window for the discovery of more objects in the category of ICS halos from Galactic sources that is becoming a new class of sources in VHE astrophysics.

\pagebreak
 
\end{document}